\documentclass{SciPost}

\hypersetup{
    colorlinks,
    linkcolor={red!50!black},
    citecolor={blue!50!black},
    urlcolor={blue!80!black}
}

\usepackage[bitstream-charter]{mathdesign}
\usepackage{graphicx}
\usepackage{subcaption}
\usepackage{amsmath}
\usepackage{array}
\usepackage{multirow}
\DeclareSymbolFont{usualmathcal}{OMS}{cmsy}{m}{n}
\DeclareSymbolFontAlphabet{\mathcal}{usualmathcal}

\fancypagestyle{SPstyle}{
\fancyhf{}
\lhead{\colorbox{scipostblue}{\bf \color{white} ~SciPost Physics Proceedings }}
\rhead{{\bf \color{scipostdeepblue} ~Submission }}

\fancyfoot[C]{\textbf{\thepage}}
}

\begin{document}

\pagestyle{SPstyle}

\begin{center}{\Large \textbf{\color{scipostdeepblue}{
Towards two-loop QCD corrections to $ \mathbf{pp \to t \bar{t} j}$ \\
}}}\end{center}

\begin{center}\textbf{
Colomba Brancaccio
}\end{center}

\begin{center}
Dipartimento di Fisica and Arnold-Regge Center, Università di Torino, and INFN, Sezione di Torino,
Via P. Giuria 1, I-10125 Torino, Italy
\\[\baselineskip]
\href{mailto:colomba.brancaccio@unito.it}{\small colomba.brancaccio@unito.it}
\end{center}

\definecolor{palegray}{gray}{0.95}
\begin{center}
\colorbox{palegray}{
  \begin{tabular}{rr}
  \begin{minipage}{0.36\textwidth}
    \includegraphics[width=60mm,height=1.5cm]{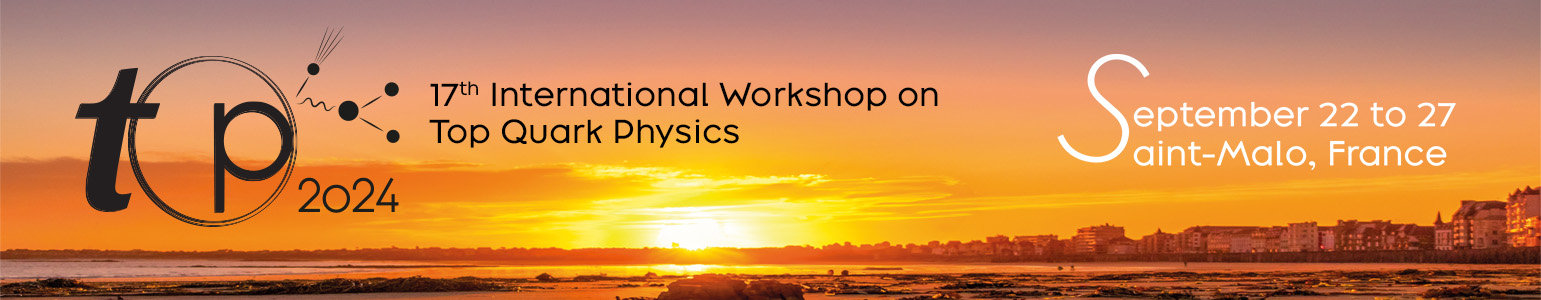}
  \end{minipage}
  &
  \begin{minipage}{0.55\textwidth}
    \begin{center} \hspace{5pt}
    {\it The 17th International Workshop on\\ Top Quark Physics (TOP2024)} \\
    {\it Saint-Malo, France, 22-27 September 2024
    }
    \doi{10.21468/SciPostPhysProc.?}\\
    \end{center}
  \end{minipage}
\end{tabular}
}
\end{center}

\section*{\color{scipostdeepblue}{Abstract}}
\textbf{\boldmath{%
I discuss the status of the computation of the two-loop QCD corrections to top-quark pair production associated with a jet at hadron colliders. This amplitude is a missing ingredient for next-to-next-to-leading order (NNLO) QCD predictions. I briefly present computational techniques to tackle the algebraic and analytic complexities of two-loop multi-scale amplitudes, in particular where massive propagators give rise to elliptic Feynman integrals. I then describe how a special function basis for the helicity amplitudes is obtained and present first numerical evaluations for the finite remainders of the $gg\to t\bar{t}g$ channel, after the infrared and ultraviolet poles have been identified analytically.
}}

\vspace{\baselineskip}

\noindent\textcolor{white!90!black}{%
\fbox{\parbox{0.975\linewidth}{%
\textcolor{white!40!black}{\begin{tabular}{lr}%
  \begin{minipage}{0.6\textwidth}%
    {\small Copyright attribution to authors. \newline
    This work is a submission to SciPost Phys. Proc. \newline
    License information to appear upon publication. \newline
    Publication information to appear upon publication.}
  \end{minipage} & \begin{minipage}{0.4\textwidth}
    {\small Received Date \newline Accepted Date \newline Published Date}%
  \end{minipage}
\end{tabular}}
}}
}


\begin{figure}[h!]
	\centering
	\includegraphics[scale=0.22]{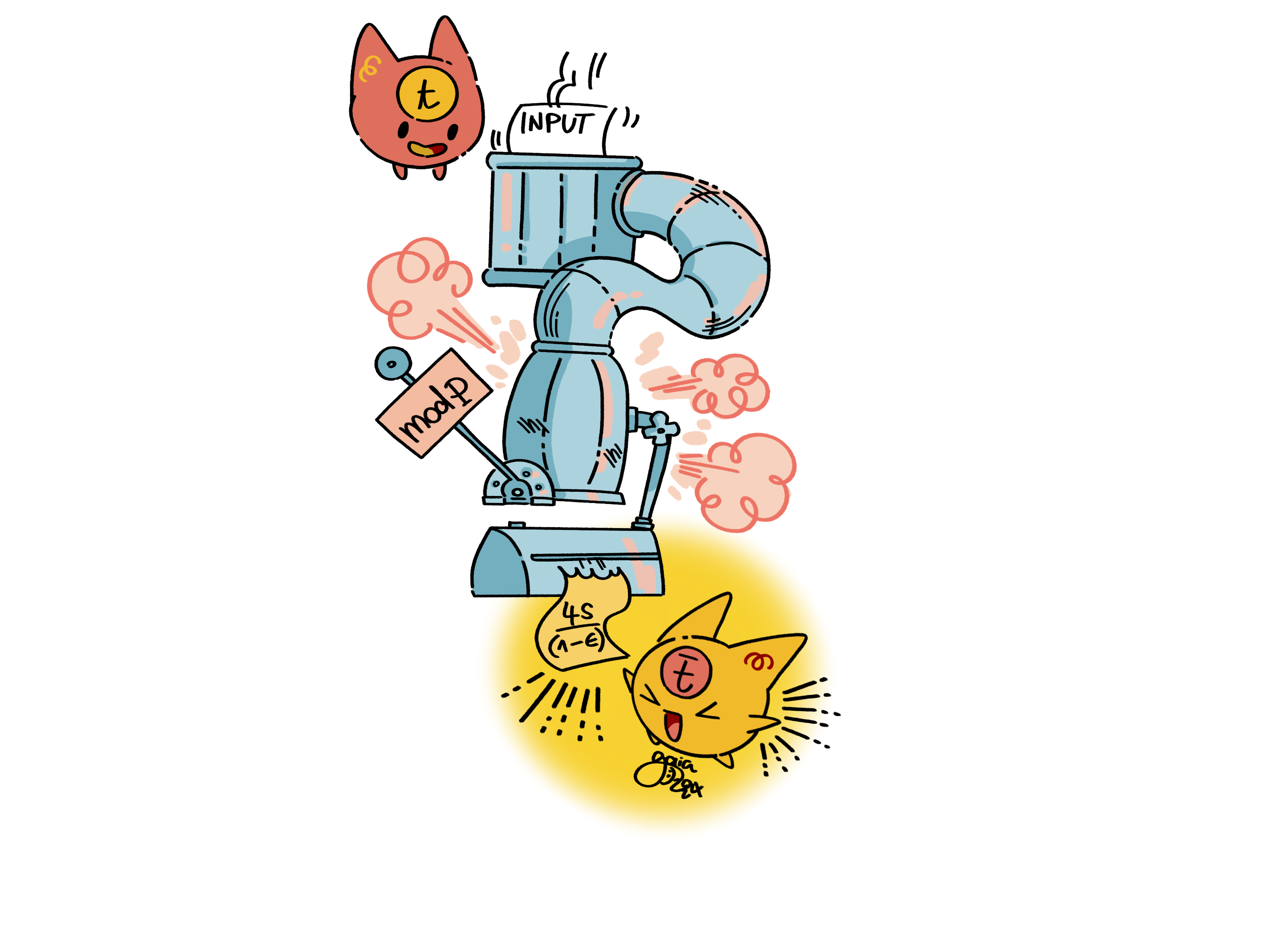}
\end{figure}

\newpage

\section{Introduction}
\label{sec:intro}
The production of a top-quark pair in association with a jet ($pp \to t\bar{t}j$) has the largest cross section among all associated production processes. As a result, this process plays a crucial role in estimating standard model backgrounds and searching for beyond standard model physics. Both the ATLAS and CMS collaborations have extensively studied this process\cite{PhysRevD.95.092001, 2018, 2020, cmscollaboration2024differentialcrosssectionmeasurements}, and more precise theoretical predictions are in high demand. In particular, the $pp \to t\bar{t}j$ normalized differential distribution with respect to the invariant mass of the final state is highly sensitive to the top quark mass\cite{alioli2013new} and can be used to infer its value.

Next-to-leading-order (NLO) QCD corrections for this process have been available for many years\cite{PhysRevLett.98.262002, dittmaier2009hadronic}. More precise predictions include full off-shell top decay effects as well as interface with parton showers\cite{MELNIKOV2010129, Alioli_2012, czakon2015matching, bevilacqua2016top, bevilacqua2016off}, and mixed NLO QCD and electroweak corrections\cite{G_tschow_2018}. In contrast, NNLO QCD corrections to $pp \to t\bar{t}j$ remain unavailable due to the high complexity of the two-loop amplitude computation.

Significant progress has been made in calculating $2 \to 3$ two-loop scattering amplitudes. This covers processes with massless external particles, for which recent examples include Refs.~\cite{badger2023isolated, Abreu:2023bdp, agarwal2024five, de2024double1, de2024double2}, as well as processes with a single external massive particle, such as in Ref.~\cite{Abreu:2021asb, Badger:2022ncb, badger2024twoloopamplitudesmathcaloalphas2corrections}. However, the computation of the $pp \to t\bar{t}j$ two-loop amplitudes is considerably more complex due to the presence of internal massive propagators in the Feynman diagrams contributing to the process (see also \cite{Agarwal:2024jyq}).

Initial steps toward addressing this challenge have been taken in Refs.~\cite{Badger:2022mrb, Badger:2022hno, Badger:2024fgb}. In the following, I present progress towards the computation of the two-loop helicity amplitudes for $pp \to t\bar{t}j$ production in the leading colour limit. I will begin by describing the amplitude workflow in Section~\ref{sec:workflow}, followed by presenting the first results in Section~\ref{sec:results}, and concluding with a summary in Section~\ref{sec:conclusion}.

\section{Framework for the scattering amplitude computation}
\label{sec:workflow}

To compute the two-loop amplitude for the process $gg \to t \bar{t} g$, we begin by generating all Feynman diagrams using \texttt{QGRAF}\cite{nogueira1993automatic}. We then proceed with the colour decomposition. Given the complexity of the amplitude, we consider the leading contribution in the number of colours. This approximation is expected to give a good estimation of the double virtual contribution, when its size is smaller compared to that of the real radiation contribution.

We compute the amplitude at leading colour using the spinor-helicity formalism\cite{Badger:2023eqz}. For massive particles, helicity states are well-defined only when a reference direction is specified. This formalism is especially useful for top-quark production, as it enables the straightforward inclusion of top-quark decays within the narrow-width approximation.

\begin{figure}[ht]
    \centering
    \begin{subfigure}[b]{0.3\textwidth}
        \centering
        \includegraphics[width=\textwidth]{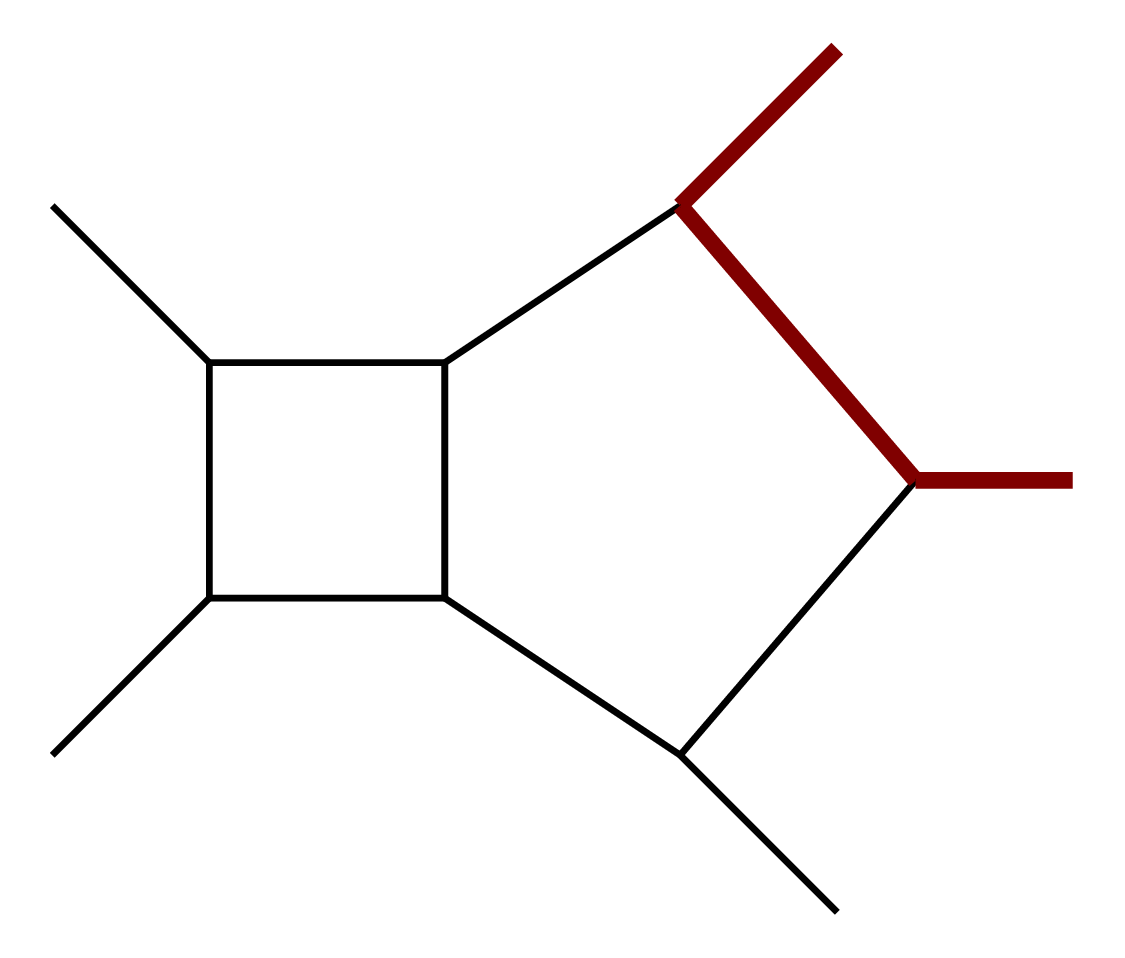}
        \caption{Topology PBA}
        \label{fig:PBA}
    \end{subfigure}
    \hfill
    \begin{subfigure}[b]{0.3\textwidth}
        \centering
        \includegraphics[width=\textwidth]{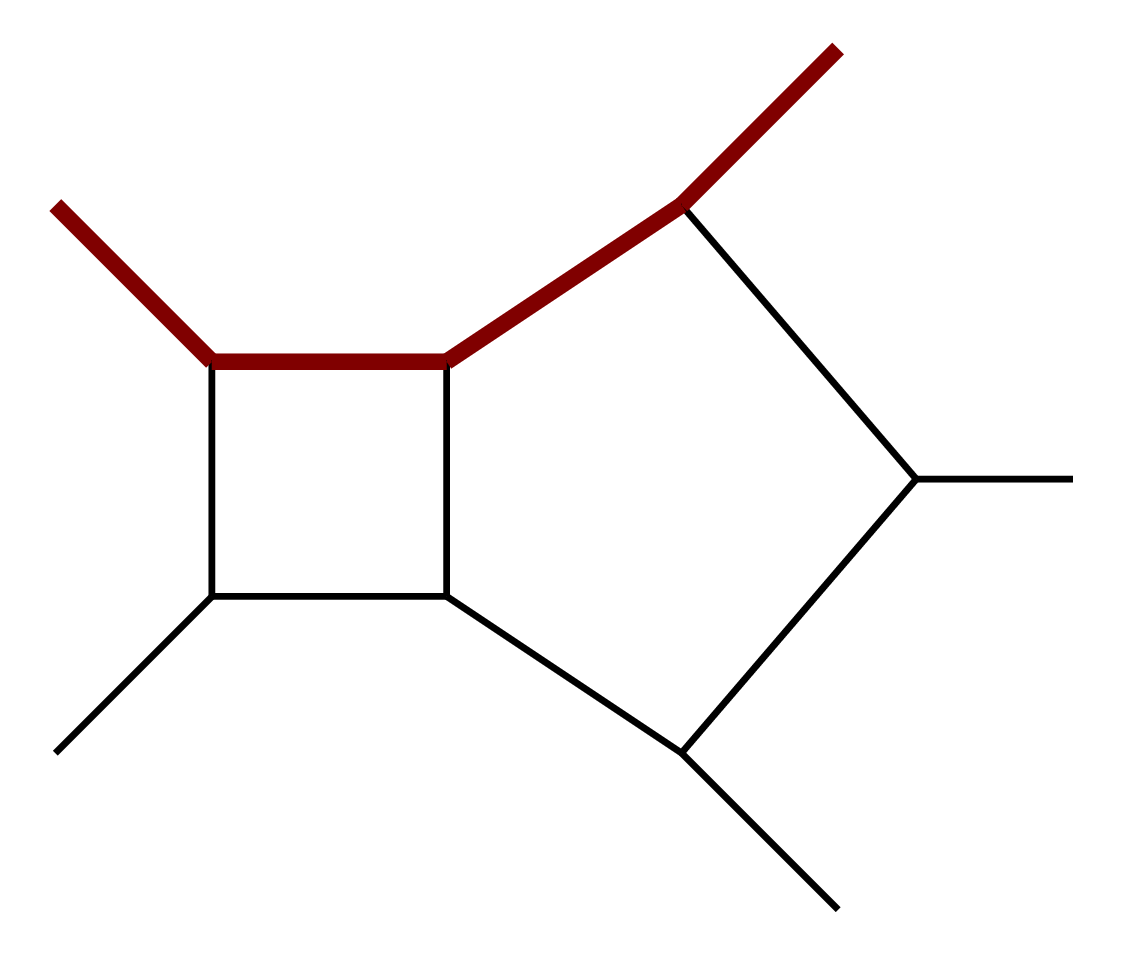}
        \caption{Topology PBB}
        \label{fig:PBB}
    \end{subfigure}
    \hfill
    \begin{subfigure}[b]{0.3\textwidth}
        \centering
        \includegraphics[width=\textwidth]{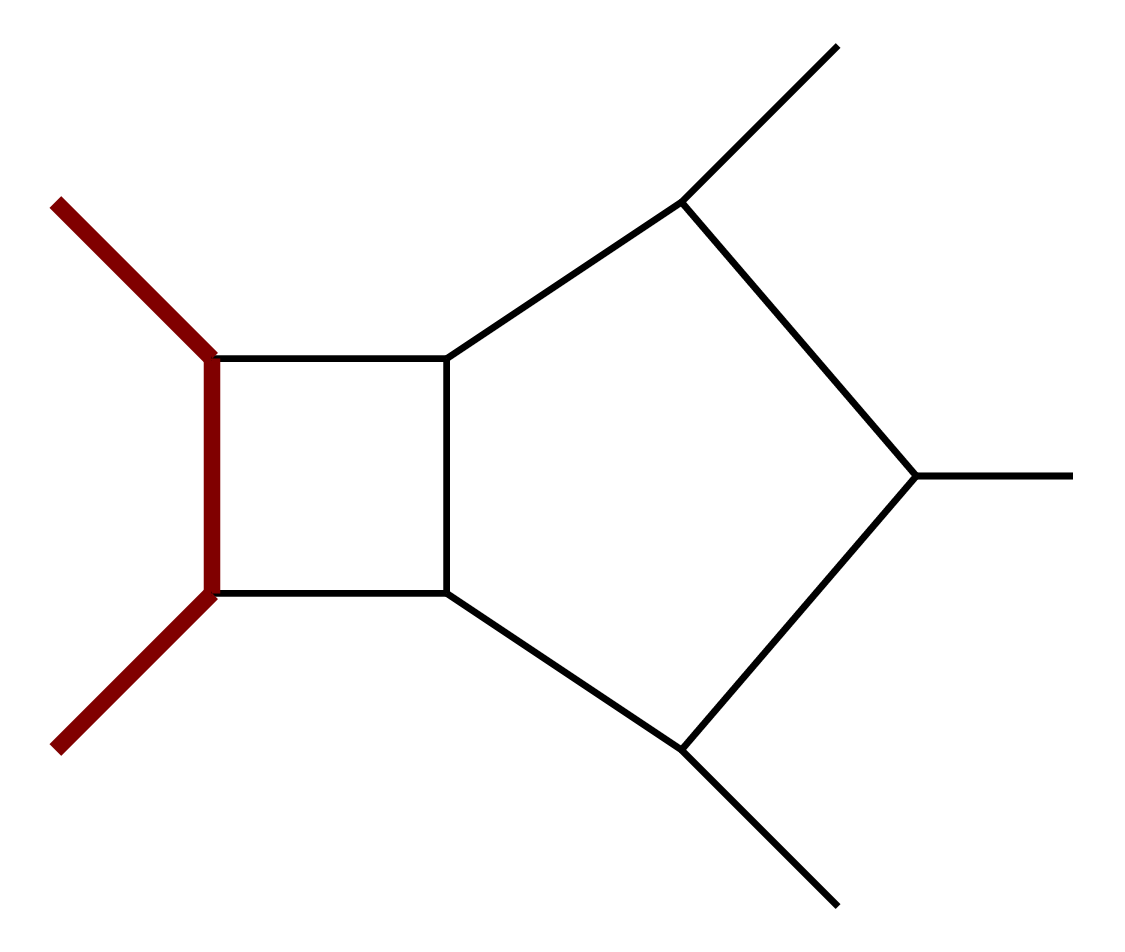}
        \caption{Topology PBC}
        \label{fig:PBC}
    \end{subfigure}
    
    \vspace{1em} 

    \begin{subfigure}[b]{0.3\textwidth}
        \centering
        \includegraphics[width=\textwidth]{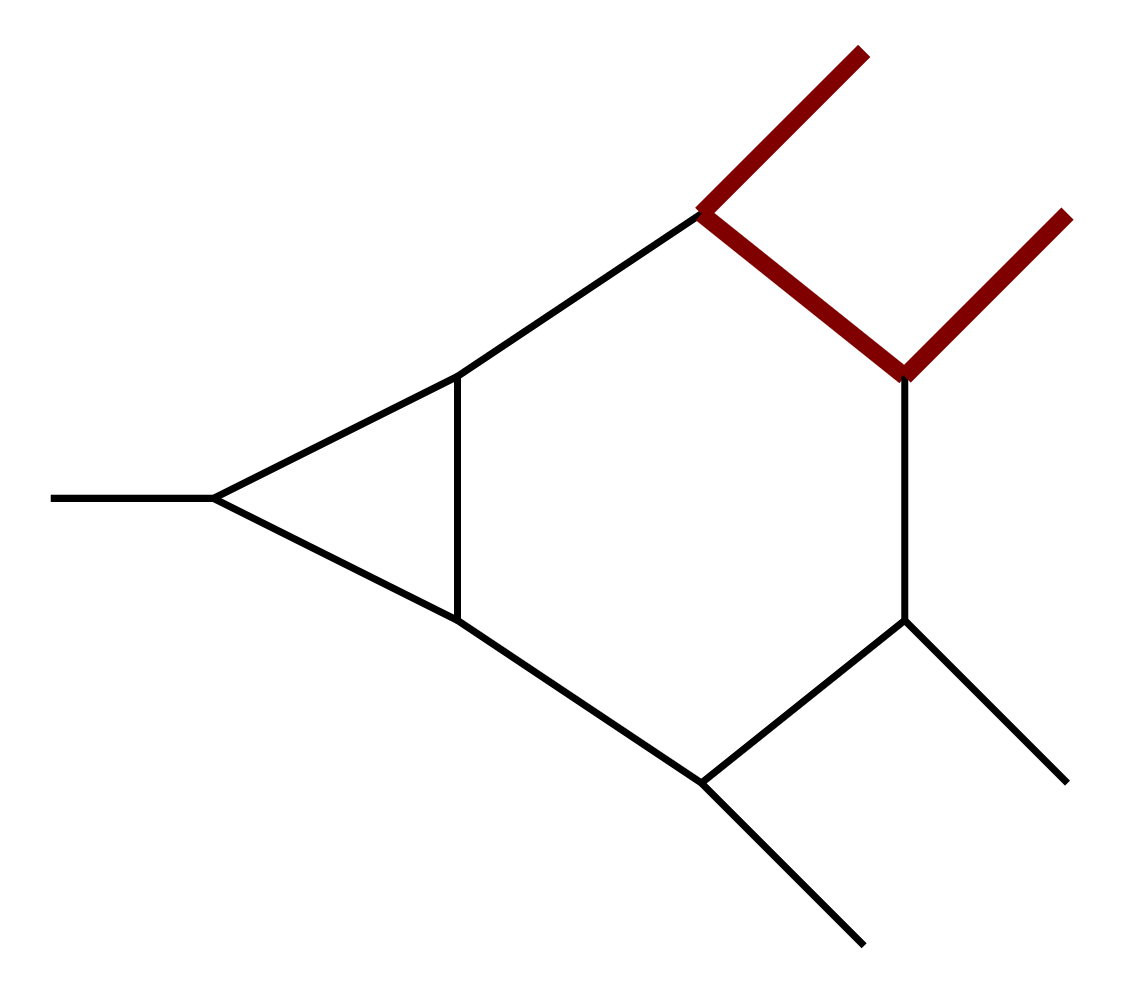}
        \caption{Topology HTA}
        \label{fig:HTA}
    \end{subfigure}
    \hfill
    \begin{subfigure}[b]{0.3\textwidth}
        \centering
        \includegraphics[width=\textwidth]{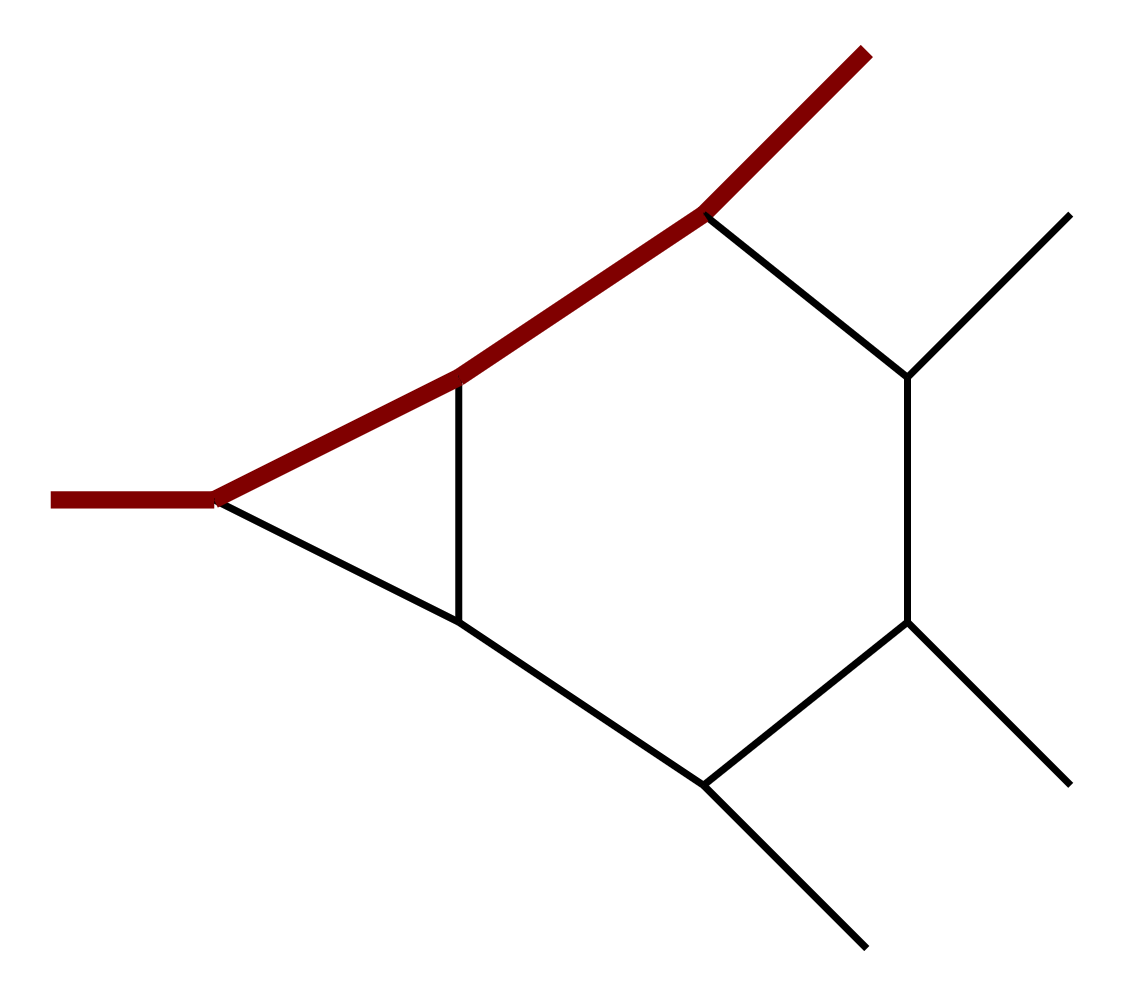}
        \caption{Topology HTB}
        \label{fig:HTB}
    \end{subfigure}
    \hfill
    \begin{subfigure}[b]{0.3\textwidth}
        \centering
        \includegraphics[width=\textwidth]{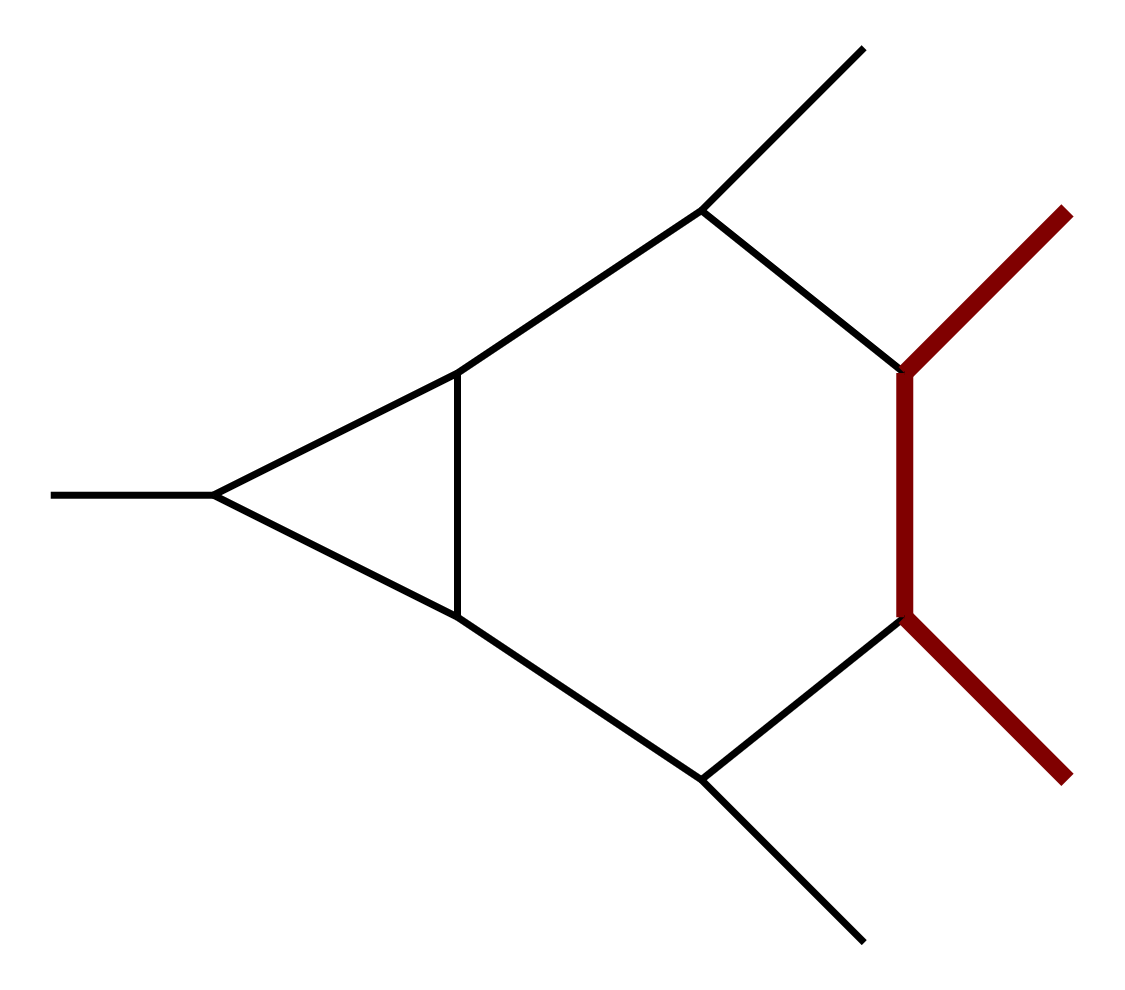}
        \caption{Topology HTC}
        \label{fig:HTC}
    \end{subfigure}

    \caption{The integral families contributing to the two-loop $pp \to t \bar{t} j$ amplitude at leading colour. Black lines represent massless particles, while thick red lines denote massive particles. All master integrals of the hexagon-triangle topologies, displayed at the bottom of the figure, can be mapped to master integrals of the three pentagon-box topologies, shown at the top.}
    \label{fig:topologies}
\end{figure}

The helicity amplitude expressions are written as linear combinations of scalar Feynman integrals. All these Feynman integrals, calculated in dimensional regularization with $d=4-2\epsilon$ spacetime dimensions, can be expressed in terms of a smaller set of linearly independent integrals, known as master integrals. The master integrals appearing at leading colour belong to six integral families, shown in Fig.~\ref{fig:topologies}. However, the master integrals of the hexagon-triangle topologies (Figs.~\ref{fig:HTA}--\ref{fig:HTC}) can be mapped to the three pentagon-box topologies (Figs.~\ref{fig:PBA}--\ref{fig:PBC}), which were recently computed in Refs.~\cite{Badger:2022hno, Badger:2024fgb}. The reduction to master integrals is achieved by solving the integration-by-parts identities\cite{Tkachov:1981wb, Chetyrkin:1981qh} generated with NeatIBP\cite{Wu:2023upw}. Expanding around $\epsilon=0$, the amplitude $A(\vec{x},\epsilon)$ takes the form
\begin{equation}
	 A(\vec{x},\epsilon)= 
	 \sum_i \sum_{k=-4}^0 \epsilon^k r_{ki}(\vec{x}) \, \text{F}_i(\vec{x})
	 + \mathcal{O}(\epsilon),
\label{eq:ampl}
\end{equation}
where $\vec{x}$ can be  a set of momentum twistor variables or Mandelstam invariants, ${F_i}$ denotes a set of special functions and $r_{ki}$ represent their coefficients, which are rational functions of the variables $\vec{x}$. Eq.~(\ref{eq:ampl}) highlights the sources of complexity inherent in these computations: algebraic and analytic.

The algebraic complexity arises from the fact that intermediate steps leading to the form shown in Eq.~(\ref{eq:ampl}) can produce extremely large expressions. The enormous size of these expressions can be streamlined by replacing symbolic operations with numerical evaluations of the rational functions. In particular, T. Peraro pioneered a method\cite{peraro2016scattering} that replaces symbolic operations with numerical evaluations over finite fields, i.e. fields of integers modulo a prime number. Using this approach and exploiting the \texttt{FiniteFlow} framework\cite{peraro2019finiteflow}, we perform intermediate computations numerically, and then reconstruct the rational coefficients of the special functions from sufficiently many numerical samples.

To obtain the amplitude expressions and perform their numerical evaluation, it is also necessary to account for special functions arising from loop integrals, introducing analytic complexity to the problem. In the case of top-quark pair production in association with a jet, the class of functions appearing in the amplitude includes not only multiple polylogarithms, which offer stable and efficient numerical evaluation, but also elliptic functions\cite{Bourjaily:2022bwx}. We extend the strategy outlined in Refs.~\cite{Gehrmann:2018yef, Chicherin:2020oor, Chicherin:2021dyp, Abreu:2023rco} to identify a set of special functions such that most of the functions are of the polylogarithmic type and algebraically independent, while the few elliptic functions contribute to the amplitudes starting from order $\epsilon^0$. This enables the analytical identification of ultraviolet (UV) and infrared (IR) poles, and leads to a dramatic simplification of the expressions. 


\section{First results for the $\mathbf{gg \to t \bar{t} g}$ two-loop amplitude at leading colour}
\label{sec:results}

We developed a framework to compute numerically the finite remainder of two-loop helicity amplitudes for top-quark pair production in association with a jet. In particular, we focus on the gluon channel, as it is the most difficult production channel to calculate.

We evaluated the two-loop helicity amplitudes for the process
\begin{equation}
	g(p_4) g(p_5) \to \bar{t}(p_1) t(p_2) g(p_3),
\end{equation}
where \(p_i\) is the external momentum. All particles are considered on-shell, with \(m_t\) representing the top-quark mass. The notation for kinematics used here is as follows:
\begin{equation} \label{eq:kinematics}
	p_1^2 = p_2^2 = m_t^2, \quad p_3^2 = p_4^2 = p_5^2 = 0, 
	\quad d_{ij} = p_i \cdot p_j.
\end{equation}
At leading colour, the amplitude for the gluonic channel takes the form
\begin{align}
	A^{(L)}(1_{\bar{t}},\, & 2_t, 3_g, 4_g, 5_g) = 
	g_s^{3+2L} N_{\epsilon}^{L} \Big\{  \nonumber\\&
	\sum_{\sigma} \left(t^{a_{\sigma(3)}} t^{a_{\sigma(4)}} t^{a_{\sigma(5)}}
	\right)^{\bar{i_1}}_{i_2} 
	A^{(L)}_{LC} (1_{\bar{t}},\, 2_t, \sigma(3)_g, \sigma(4)_g, \sigma(5)_g)
	\Big\},
\end{align} 
where \(L\) is the loop order, $g_s$ is strong coupling and $N_{\epsilon} = \dfrac{e^{\epsilon \gamma_E} \Gamma^2(1-\epsilon) \Gamma(1+\epsilon)}{(4 \pi)^{2-\epsilon} \Gamma(1-2\epsilon)}$ the overall normalisation. Moreover, $\sigma$ denote the cyclic gluon permutations, and \(\left( t^a \right)^{\bar{j}}_i\) are fundamental generators of the \(SU(N_c)\) colour group, with \( a = 1, \dots, 8 \) indexing the adjoint representation and \( i, \bar{j} = 1, 2, 3 \) indexing the fundamental and anti-fundamental representations, respectively.

To handle the helicity states of massive fermions and retain the decay information within the narrow-width approximation, we construct the amplitude using the following spin structure basis,
\begin{align}
	A_{LC}^{(L)}(1_{\bar{t}}^+,2_t^+,
	3^{h_3}&,4^{h_4},5^{h_5};n_t,n_{\bar{t}}) =
  m_t \Phi(3^{h_3},4^{h_4},5^{h_5})\nonumber\\&
  \sum_{i=1}^4 \Theta_i(1,2;n_t,n_{\bar{t}}) \, 
  A_{LC}^{(L),[i]}(1_{\bar{t}}^+,2_t^+,3^{h_3},4^{h_4},5^{h_5}).
\label{eq:helampl}
\end{align}
In Eq.~(\ref{eq:helampl}), \(h_i\) represents the helicity configuration of each massless final-state particle, and \(n_t\) (\(n_{\bar{t}}\)) is an arbitrary reference vector defining the massive spinors. Because \(n_t\) and \(n_{\bar{t}}\) are arbitrarily chosen, positive and negative helicity states are related. Thus, we only consider the \(++\) helicity configuration for the top-quark pair. The phase factor \(\Phi\) encodes the helicity information of the massless particles, while the \(\Theta\) factors capture the spin structure of the top-quark pair relative to the reference vectors \(n_t\) and \(n_{\bar{t}}\). Further details on this formalism and explicit definitions of \(\Phi\) and \(\Theta\) can be found in Refs.~\cite{Kleiss:1985yh,Badger:2021owl,Badger:2022mrb}.

We computed the mass counterterm enabling us to perform the Ward identity check and confirm that our results are gauge invariant. We further performed an analytical check of the poles in the amplitude, comparing against UV renormalization and IR subtraction\cite{Becher:2009kw, Becher_2009}. The analytic pole check is a non-trivial task made possible by introducing a basis of special functions in which elliptic functions appear only starting from order $\epsilon^0$. This approach allowed us to directly extract the finite remainder in the ’t Hooft–Veltman scheme.

In Table~\ref{tab:results}, we present results for the finite remainder of the helicity configuration \(h_i  = + \), with \(i  = 3,4,5 \), at a few benchmark phase-space points. The decay directions of the top quarks are fixed along a non-physical direction, with \(n_t = n_{\bar{t}} = p_3\). These values were obtained by computing the special functions appearing in the amplitude via differential equations, using the generalized series expansion method\cite{Moriello:2019yhu}, employing the \texttt{DiffExp} software\cite{Hidding:2020ytt}.

\newpage

\begin{table}[h!]
\centering
\renewcommand{\arraystretch}{1.4} 
\setlength{\tabcolsep}{12pt} 
\begin{tabular}{|>{\centering\arraybackslash}p{8cm}|>{\centering\arraybackslash}p{5cm}|} 
 \hline
   Phase-space points & \textbf{$A_{LC}^{(2)}(+++++; n_t n_{\bar{t}}) \, [\mathrm{GeV}^{-2}]$} \\
 \hline
 \hline
 $\begin{array}{c}
    d_{12} \to 0.1074, \, d_{23} \to 0.2719, \, d_{34} \to -0.1563, \\
    d_{45} \to 0.5001, \, d_{15} \to -0.03196, \, m_t^2 \to 0.02502
 \end{array}$ & $19.03 - 3.108 \, i$ \\
 \hline
 $\begin{array}{c}
    d_{12} \to 0.3915, \, d_{23} \to 0.06997, \, d_{34} \to -0.06034, \\
    d_{45} \to 0.5002, \, d_{15} \to -0.1293, \, m_t^2 \to 0.02499
 \end{array}$ & $0.07061 - 0.006497 \, i$ \\
 \hline
 $\begin{array}{c}
    d_{12} \to 0.2167, \, d_{23} \to 0.02186, \, d_{34} \to -0.01149, \\
    d_{45} \to 0.5007, \, d_{15} \to -0.04709, \, m_t^2 \to 0.02502
 \end{array}$ & $-29.22 - 27.54 \, i$ \\
 \hline
 $\begin{array}{c}
    d_{12} \to 0.2986, \, d_{23} \to 0.1599, \, d_{34} \to -0.05978, \\
    d_{45} \to 0.4998, \, d_{15} \to -0.2899, \, m_t^2 \to 0.02500
 \end{array}$ & $-0.9728 + 0.8636 \, i$ \\
 \hline
 $\begin{array}{c}
    d_{12} \to 0.2882, \, d_{23} \to 0.04770, \, d_{34} \to -0.1080, \\
    d_{45} \to 0.5000, \, d_{15} \to -0.1583, \, m_t^2 \to 0.02502
 \end{array}$ & $-0.4041 - 0.5316 \, i$ \\
 \hline
\end{tabular}
\caption{Numerical evaluations of the leading colour two-loop helicity amplitude \(A_{LC}^{(2)}(+++++; n_t n_{\bar{t}})\) for different phase-space points in the physical region. In this table, \(d_{ij}\) denotes the kinematic invariants and \(m_t\) the top-quark mass. The kinematic invariants are normalised to $2 \, d_{45}$ and the sign of the imaginary part of $\text{tr}_5$ is chosen to be positive; their rationalised values are available on request. The reference vectors \(n_t\) and \(n_{\bar{t}}\) are chosen in the unphysical direction \(p_3\) and are used to define the spinors for the top quark and the anti-top quark.}
\label{tab:results}
\end{table}

\section{Conclusions and outlook}
\label{sec:conclusion}

We reported progress in computing two-loop helicity amplitudes for top-quark pair production in association with a jet, specifically focusing on the $gg \to t \bar{t} g$ process, which is the most challenging production channel to compute. This computation represents a crucial step toward achieving NNLO QCD corrections for this high-priority process at hadron colliders.

We developed a framework for evaluating the finite remainder of two-loop helicity amplitudes for the process under consideration, addressing the extreme algebraic and analytic complexities inherent in multi-scale processes that involve elliptic functions. We checked gauge invariance validating the Ward identity, and analytically identified the UV and IR poles using a basis of special functions, where the elliptic sector contributes only at order $\epsilon^0$. These poles were then checked against UV renormalization and IR subtraction, enabling us to extract the finite remainder of the helicity amplitudes. As an outlook, we aim to deliver phenomenologically viable results and explore the feasibility of a full analytical reconstruction.

\section*{Acknowledgements}
I want to thank Simon Badger, Matteo Becchetti, Heribertus Bayu Hartanto and Simone Zoia for their collaboration and commitment to this work. Thanks to Gaia Fontana for the wonderful drawing on the front page. This project has been supported by the European Union’s Horizon Europe research and innovation programme under the Marie Sklodowska-Curie grant agreement No. 101105486, and the ERC grant No. 101040760 FFHiggsTop. My work is funded from the Italian Ministry of Universities and Research through FARE grant R207777C4R. 

\bibliography{CB_TOP24.bib}

\end{document}